\documentclass[
      final            % use final for the camera ready runs
  ]{aipproc}
\usepackage{verbatim}
\layoutstyle{8x11single}
\begin{document}

\title{Von Neumann-L\"uders projection, and its applicability to EPR experiments}

\classification{02.50.Cw, 03.65.Ca, 03.65.Ta, 03.65.Ud}
\keywords{von Neumann-L\"uders projection, conditional
preparation, EPR}

\author{Willem M. de Muynck}{
address={Theoretical Physics, Eindhoven University of Technology, Eindhoven, the
Netherlands}
}

\begin{abstract}
Notwithstanding it is well known that von Neumann's projection
postulate is inapplicable to most realistic measurement
procedures, it keeps haunting the foundations of quantum
mechanics. In particular its applicability to EPR experiments is
often assumed. In the present contribution this problem is
considered from the point of view of a quantum mechanical theory
of measurement, allowing a treatment of projection in EPR
experiments as a special case of conditional preparation. The
conditions are spelled out under which the postulate may be
applicable.
\end{abstract}

\maketitle

%%%%%%%%%%%%%%%%%%%%%%%%%%%%%%%%%%%%%%%%%%%%
%% MAINMATTER
%%%%%%%%%%%%%%%%%%%%%%%%%%%%%%%%%%%%%%%%%%%%
\begin{comment}
Ref?
R. Laura, L. Vanni
Int J Th Ph DOI 10.1007/s10773-008-9672-7
Conditional Probabilities and Collapse in Quantum Measurements
\end{comment}

\section{1. Von Neumann-L\"uders projection, measurement or preparation?}\label{sec1}

Von Neumann's projection postulate tells us that, if a measurement
is performed of observable ${\bf A} = \sum_m a_m {\bf P}_m$, ${\bf
P}_m = |a_m\rangle\langle a_m|$, and the (non-degenerate)
eigenvalue $a_m$ is obtained as a measurement result, then the
state vector makes a transition
\begin{equation}
|\psi\rangle \rightarrow |a_m\rangle
\label{1}
\end{equation}
from the initial state $|\psi\rangle = \sum_m c_m |a_m\rangle$ to
the corresponding eigenvector $|a_m\rangle$. In case of degeneracy
this is generalized to the L\"uders projection
\begin{equation}
\label{2}
|\psi\rangle \rightarrow \frac{{\bf P}_m|\psi\rangle}{\parallel\! {\bf P}_m|\psi\rangle\!\parallel},
\end{equation}
${\bf P}_m =\sum_i |a_{mi}\rangle\langle a_{mi}|$ the projection
operator on the subspace of Hilbert space spanned by the
eigenvectors of $\bf A$ for which $a_{mi}=a_{m}$.

As a reason for assuming the projection postulate von Neumann
(\cite{vN32} section III.3) refers to the Compton-Simon experiment
\cite{ComSim25} in which a $\gamma$ photon is scattered by an
electron (initially having linear momenta ${\bf p}_{ph}$ and
 ${\bf p}_e$, respectively), yielding the state
\begin{equation}
|\psi_{\rm CS}\rangle = \int d{\bf p}'_e c({\bf p}'_e)
 |{\bf p}'_e\rangle|{\bf p}'_{ph} \rangle,\;{\bf p}'_{ph} = {\bf p}_{ph}
+ {\bf p}_e - {\bf p}'_e. \label{3}
\end{equation}
In the experiment ${\bf p}_{ph}$ and ${\bf p}_e$ are assumed to be
known (actually ${\bf p}_e$ is supposed to be negligibly small
compared to the momenta of the $\gamma$ ray photons). Then, by
performing in the state (\ref{3}) a measurement of electron
momentum and applying conservation of linear momentum it is
possible to obtain from the measurement result ${\bf p}'_e$ the
corresponding photon momentum according to ${\bf p}'_{ph} = {\bf
p}_{ph} - {\bf p}'_e$. It is concluded that the photon must have
this momentum value with certainty, and that the determination of
${\bf p}'_e$ must be accompanied by a transition of the state of
the photon to the corresponding momentum eigenstate $|{\bf
p}'_{ph}\rangle$. This is corroborated by consecutive experiments
performed using the scattered photons.

Scattering experiments like the Compton-Simon experiment have been
the main tools for testing quantum mechanics during the initial
time of its development. As a consequence these experiments could
acquire a paradigmatic status with respect to general quantum
mechanical measurement, requiring von Neumann projection to be
valid for \emph{all} measurements. This is impossible, however.
Thus, it is evident that (\ref{1}) cannot be applied in a rigorous
way to observables having continuous spectra. More importantly,
the Compton-Simon experiment has lost its paradigmatic status
since many present-day measurement procedures simply do \emph{not}
satisfy (\ref{1}) or (\ref{2}). This holds true even for the
Stern-Gerlach experiment, which is often presented as satisfying
the postulate, the outgoing beams being associated with
eigenvectors of the measured angular momentum observable. For two
reasons this cannot be true, however. First, `the atom being in an
outgoing beam' does not constitute a measurement. It is a
measurement only if it is ascertained which of the beams the atom
actually is in. For this reason detectors must be placed in the
beams, monitoring the presence of the atom. Even if in the
Stern-Gerlach experiment by the interaction of the atom with the
magnetic field the outgoing states were prepared as eigenvectors
of the angular momentum observable, then they can hardly be
expected to remain such eigenstates while interacting with a
detector which is not designed to leave angular momentum
uninfluenced. Second, not even the assumption is warranted that
the interaction of the atom with the magnetic field prepares the
states of the beams as eigenvectors of the angular momentum
observable. It is by now well known (\cite{MadM93}, \cite{dM2002}
section~8.3.2) --but, unfortunately, widely ignored-- that
inhomogeneity of the Stern-Gerlach magnetic field is incompatible
with the conservation of angular momentum necessary for the
assumption to be satisfied. A requirement to satisfy von Neumann
projection could be fulfilled only if the magnetic field were
homogeneous. Since inhomogeneity of the magnetic field is a
necessary condition for the measurement to be effective (be it not
without a certain inaccuracy \cite{MadM93}) it turns out that the
Stern-Gerlach experiment is an appropriate procedure for the
measurement of angular momentum only as a consequence of
\emph{not} satisfying von Neumann projection. This example can be
supplemented with many others. For this reason it seems wise not
to rely on (\ref{1}) or (\ref{2}) when discussing general
properties of quantum measurement. That the projection postulate
is controversial has been argued before, for instance, by Wigner
\cite{Wig63}, Ballentine \cite{Bal90a} and by Cini and
L\'evy-Leblond \cite{CiLL90}.

As a consequence of taking into account a too restricted class of
measurements, viz. scattering experiments like the Compton-Simon
and the Stern-Gerlach ones, it seemed that \emph{preparing} the
object in some final state was the essence of measurement. The
conception of a measurement as a preparation of a microscopic
object in a particular final state has become one of the
characteristics of the Copenhagen interpretation (Heisenberg
\cite{Heis30} section~II.2). Nevertheless, it has been realized
already a long time ago that von Neumann projection is
characterizing only a subset of measurement procedures, referred
to as `measurements of the first kind'. Experiments \emph{not}
preparing the microscopic object in an eigenstate of the measured
observable are often referred to as `measurements of the second
kind'.

Confusion of the preparative and determinative aspects of
measurement is at the basis of von Neumann's projection postulate
(de Muynck \cite{dM2000}). It could arise because no serious
theoretical attention was paid to the actual measurement process
involved in the interaction of the microscopic object with a
(macroscopic) measuring instrument. Admittedly, von Neumann
(\cite{vN32} section VI.3) did consider such interaction. However,
his treatment was meant to justify his projection postulate by
demonstrating that the latter is consistent with measurement as a
quantum mechanical interaction process, rather than to critically
examen the feasibility of such processes. Only after it was
realized that in measurement procedures the important feature is
the preparation of the \emph{measuring instrument} (yielding a final
pointer position $m$, cf. figure~\ref{fig1}) rather than the
preparation of the microscopic object (represented in figure~\ref{fig1}
by the states $|\psi_m\rangle$), was it possible to evaluate
the role of von Neumann-L\"uders projection in quantum mechanical
measurement.

\section{2. Quantum mechanical theory of measurement, conditional preparation, and von Neumann
projection}
\label{sec2} Von Neumann tried to justify the transition (\ref{1})
by assuming the interaction between object and measuring
instrument to induce a transition of the state vector of the
system `object+measuring instrument' according to
\begin{equation}
|\Psi\rangle =\sum_m c_m |a_m\rangle |\theta\rangle \rightarrow
 |\Psi_f \rangle = \sum_{m}c_m |a_m \rangle  |\theta_m \rangle,
\label{3a}
\end{equation}
in which $|\theta\rangle$ is the initial state of the measuring
instrument, and $|\theta_m \rangle$ are its final pointer states.
These latter states are assumed to be mutually orthogonal
eigenvectors of the pointer observable $\Theta$. Finding final
pointer position $m$ is interpreted as obtaining measurement
result $a_m$. If (\ref{3a}) is correct, then (\ref{1}) can be
justified by performing in the final state $|\Psi_f\rangle$ a
joint measurement of the pointer observable $\Theta$ and an
\emph{arbitrary} observable $B=\sum_n b_n |b_n\rangle\langle b_n|$
of the object, yielding the joint probability distribution of
these compatible observables according to
\begin{equation}
\label{14} p(m,b_n) = |\langle b_n|\langle
\theta_m|\Psi_f\rangle|^2 = |c_m|^2|\langle b_n|a_m\rangle|^2.
\end{equation}
From this the conditional probability $p(b_n|m)$ of measurement
result $b_n$, given the final pointer position $m$, is found as
\begin{equation}
\label{13} p(b_n|m) = \frac{p(m,b_n)}{p(m)} = |\langle
b_n|a_m\rangle|^2.
\end{equation}
From the arbitrariness of $B$ it follows that the state of the
object, conditional on pointer position $m$, must be given by
$|a_m\rangle$, as required by (\ref{1}).

Unfortunately, the measurement scheme (\ref{3a}) refers only to
measurements of the first kind. In order to take into account the
possibility of measurements of the second kind the scheme
(\ref{3a}) should be replaced by
\begin{equation}
|\Psi\rangle =\sum_m c_m |a_m\rangle  |\theta\rangle \rightarrow
 |\Psi_f \rangle = \sum_{m}c_m |\psi_m \rangle  |\theta_m \rangle,
\label{3b}
\end{equation}
in which the vectors $|\psi_m \rangle$ are normalized but in
general not mutually orthogonal. What is the final state of the
object is determined by the interaction between object and
measuring instrument. In general this interaction disturbs the
object so as to end up, conditional on pointer position $m$, in a
state $|\psi_m \rangle$ \emph{different} from $|a_m\rangle$, even
if this was the object's initial state. Analogously to (\ref{14})
and (\ref{13}) it is found that (\ref{1}) should be replaced by
the transition
\begin{equation}
|\psi\rangle \rightarrow |\psi_m\rangle.
\label{1b}
\end{equation}
Hence, in general von Neumann's projection postulate is not
satisfied (nor is the L\"uders one).
\begin{figure}[t]
\leavevmode \centerline{\includegraphics[height=3.5cm]{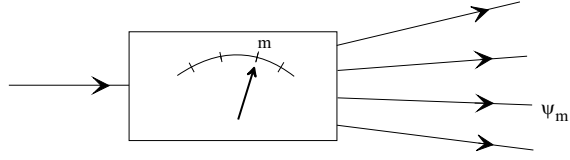} }
\caption{\em \small Determinative and preparative aspects of
measurement.} \label{fig1}
\end{figure}

The theory of measurement allows to distinguish between
determinative and preparative aspects of measurement (cf.
figure~\ref{fig1}). The determinative aspect refers to the
knowledge on the initial state to be obtained by determining the
final pointer position of a measuring instrument. The preparative
aspect refers to what has become of the object after the
measurement. These are completely different features, being
related by von Neumann's projection postulate in an untoward way.
Transitions like (\ref{1}) and (\ref{1b}) refer to
\emph{conditional preparations}, preparing the object in state
$|a_m\rangle$ or $|\psi_m\rangle$ if measurement result $m$ is
obtained. If the object is not absorbed by the measuring
instrument, then it is possible in principle to perform a
consecutive experiment while selecting outgoing objects on the
basis of pointer positions $m$, so as to be sure to have
conditionally prepared states. The idea that von Neumann-L\"uders
projection is a general property of measurement stems from a
confusion of the determinative and preparative aspects of quantum
measurement, transitions like (\ref{1}) and {\ref{2}) describing
preparation procedures rather than measurements.

\section{3. Von Neumann-L\"uders projection and EPR}
\label{sec3} It is evident that, if applicable at all to EPR
\cite{EPR}, we should consider L\"uders projection rather than von
Neumann's, since observable ${\bf A}^{(1)}$ of particle 1 should
actually be considered as an operator ${\bf A}^{(1)}
 {\bf I}^{(2)}$ on a two-particle space, which has
degenerate eigenvalues even if the eigenvalues of ${\bf A}^{(1)}$
are non-degenerate (which shall be assumed here). Taking the
initial EPR state according to
\begin{equation}\label{11}
|\psi_{\rm EPR}\rangle = \sum_m c_m |a^{(1)}_m\rangle
|a^{(2)}_m\rangle,
\end{equation}
we get
\begin{equation}
\label{4} {\bf P}_m = |a^{(1)}_m\rangle\langle
a^{(1)}_m|\sum_{m'}|a^{(2)}_{m'}\rangle\langle a^{(2)}_{m'}|=
|a^{(1)}_m\rangle\langle a^{(1)}_m| {\bf I}^{(2)},
\end{equation}
in the state $|\psi_{\rm EPR}\rangle$ yielding a L\"uders
projection (\ref{2}) according to
\begin{equation}
\label{6}
|\psi_{\rm EPR}\rangle \rightarrow |a^{(1)}_m\rangle
 |a^{(2)}_m\rangle.
\end{equation}
This result is in agreement with our experience with respect to
measurements consecutively performed on particle 2. However, it
does not take into account a possible disturbance of particle 1 by
the measurement of ${\bf A}^{(1)}$ actually carried out. Indeed,
taking such a disturbance into account, application of the theory
of measurement to EPR yields transitions (\ref{3b}) and (\ref{1b})
according to
\begin{equation}
|\Psi\rangle =\sum_m c_m |a^{(1)}_m\rangle |a^{(2)}_m\rangle
|\theta\rangle \rightarrow
 |\Psi_f \rangle = \sum_m c_m |\psi^{(1)}_m\rangle |a^{(2)}_m\rangle  |\theta_m \rangle
\label{3c}
\end{equation}
and
\begin{equation}
\label{7} |\psi_{\rm EPR}\rangle \rightarrow |\psi^{(1)}_m\rangle
 |a^{(2)}_m\rangle,
\end{equation}
respectively. Hence, it is allowed to apply von Neumann projection
to particle 2. But, as before, the measurement interaction
prevents the postulate in general to be satisfied by the particle
that has interacted with the measuring instrument.

Comparing (\ref{3}) and (\ref{11}) a close similarity can be
observed between the EPR and Compton-Simon experiments, the latter
serving von Neumann as an incentive to assume his projection
postulate (\ref{1}). In both cases it is the object \emph{not}
interacting with the measuring instrument that is assumed to be
subjected to projection, the actual measurement being carried out
on another object. In both cases the object not interacting with
the measuring instrument is not actually subjected to a
measurement, but it is \emph{prepared} in a quantum mechanical
state ($|{\bf p}'_{ph}\rangle$ and $|a^{(2)}_m\rangle$,
respectively), which can act as an initial state in a new
experiment (not necessarily a measurement) performed with the
photons c.q. particles corresponding to results ${\bf p}'_{e}$
c.q. $a^{(1)}_m$ obtained from measurements on correlated objects.
From this it is evident that, although von Neumann projection may
not be applicable to measurement, it may yield a description of
\emph{conditional preparation}. Indeed, conditional preparation is
a widely used experimental procedure for preparing an object in a
well-defined state, unfortunately often referred to as a
`measurement', for instance, as a `{\em predictive} measurement'
(Kemble \cite{Kemble} section~41). Von Neumann-L\"uders projection
is applicable only as far as the measurement does not disturb the
object.

\section{4. General theory of conditional preparation}
\label{sec4} The method of conditional preparation can easily be
extended to less simplistic, and possibly more realistic
measurement procedures than considered in sections~\textbf{2} and
\textbf{3}. Indeed, the initial states of object and measuring
instrument may be represented by density operators
 ${\bf \rho}^{(o)}$ and ${\bf \rho}^{(a)}$, respectively. Let the
measurement interaction yield as its final state
\begin{equation}\label{15}
{\bf \rho}_f = {\bf U\rho}^{(o)}{\bf \rho}^{(a)}{\bf U}^\dagger, \; {\bf U} = \exp{-i{\bf H}T},
\end{equation}
$\bf H$ the interaction Hamiltonian, $T$ the interaction time.
Then the detection probabilities are given by
\begin{equation}\label{16}
p(m) = Tr_{oa} {\bf \rho}_f {\bf E}^{(a)}_m,\; {\bf E}^{(a)}_m =
\sum_i |\theta_{mi}\rangle\langle \theta_{mi}|,
\end{equation}
in which the $|\theta_{mi}\rangle$ are the degenerate eigenvectors
of the pointer observable representing the pointer's microscopic
structure at pointer position $m$.

The conditionally prepared state can be found, as before, by
contemplating in the final state a joint measurement of
the pointer observable and an arbitrary observable of the object,
yielding joint probabilities
\begin{equation}\label{17}
p(mn)  = Tr_{oa} {\bf \rho}_f {\bf E}^{(a)}_m {\bf F}^{(o)}_n.
\end{equation}
The conditionally prepared states ${\bf \rho}^{(o)}_{fm}$ can now
be found as those object states for which the conditional
probabilities $p(n|m)$, related to $p(mn)$ and $p(m)$ according to
$p(mn)=p(n|m)p(m)$, satisfy
\begin{equation}\label{18}
p(n|m) = Tr_o{\bf \rho}^{(o)}_{fm}{\bf F}^{(o)}_n.
\end{equation}
As a consequence of the arbitrariness of $\{{\bf F}^{(o)}_n\}$ it
straightforwardly follows from (\ref{17}) and (\ref{18}) that
\begin{equation}\label{8}
{\bf \rho}^{(o)}_{fm} = \frac{Tr_a{\bf \rho}_f
 {\bf E}^{(a)}_m}{Tr_{oa} {\bf \rho}_f {\bf E}^{(a)}_m}.
\end{equation}
As demonstrated by de Muynck (\cite{dM2002} section~3.3.4) this
derivation applies even if the measurement is a generalized one in
which the probabilities (\ref{16}) correspond to a positive
operator-valued measure $\{{\bf M}_m =
 Tr_a {\bf \rho}^{(a)}{\bf U}^\dagger {\bf E}^{(a)}_m {\bf U}\}$. In particular,
 if ${\bf M}_m =\sum_{m'}\lambda_{mm'}
|a_{m'}\rangle\langle a_{m'}|,\; \lambda_{mm'}\geq 0,\; \sum_m
\lambda_{mm'} =1$, the measurement is a \emph{nonideal}
measurement of the standard observable ${\bf A}$  (de Muynck
\cite{dM2002} section~7.6.5). This is satisfied if
\begin{equation}\label{20}
Tr_{oa} {\bf U}|a_{m'}\rangle\langle a_{m''}|
  {\bf \rho}^{(a)}{\bf  U}^\dagger {\bf E}^{(a)}_m =\lambda_{mm'}\delta_{m'm''}.
\end{equation}
Such nonideal measurements are encountered in many practical
applications of quantum measurement theory (e.g. de Muynck
\cite{dM2002} chapter~8).

 The theory of conditional preparation can straightforwardly
be applied to the EPR experiment by specifying the object ($o$) to
consist of 2 particles (labeled 1 and 2, respectively) and taking
${\bf F}^{(o)}_n =
 {\bf F}^{(1)}_{n_1}{\bf F}^{(2)}_{n_2}$. Analogously to
(\ref{8}) we find as the conditionally prepared
state of the two-particle system
\begin{equation}\label{12}
{\bf \rho}^{(12)}_{fm} = \frac{Tr_a{\bf \rho}_f
 {\bf E}^{(a)}_m}{Tr_{12a} {\bf \rho}_f {\bf E}^{(a)}_m}.
\end{equation}

Let us apply this to the EPR state $|\psi_{\rm EPR}\rangle =
\sum_m c_m |a^{(1)}_m\rangle |a^{(2)}_m\rangle$ (\ref{11}), taking
the measurement to be of observable ${\bf A}^{(1)}$ of particle 1
only. Since the interaction operator $\bf U$ does not depend on
the particle 2 variables we have $[{\bf U},
|a^{(2)}_{m'}\rangle\langle a^{(2)}_{m''}|]_- = {\bf O}$. From
(\ref{12}) it then follows that
\begin{equation}\label{21}
{\bf \rho}^{(2)}_{fm} = Tr_1{\bf \rho}^{(12)}_{fm} =
\frac{1}{p(m)} \sum_{m'm''}  c_{m'}c^*_{m''}
|a^{(2)}_{m'}\rangle\langle a^{(2)}_{m''}| Tr_{1a} {\bf
U}|a^{(1)}_{m'}\rangle\langle a^{(1)}_{m''}|
 {\bf \rho}^{(a)}{\bf  U}^\dagger {\bf E}^{(a)}_m.
\end{equation}
If we take in (\ref{20}) particle 1 for object $o$, then for a
nonideal measurement of ${\bf A}^{(1)}$ we obtain from (\ref{20})
and (\ref{21})
\begin{equation}\label{22}
 {\bf \rho}^{(2)}_{fm} =
\sum_{m'}\frac{\lambda_{mm'}|c_{m'}|^2}{\sum_{m''}\lambda_{mm''}|c_{m''}|^2}|a^{(2)}_{m'}\rangle\langle
a^{(2)}_{m'}|,
\end{equation}
demonstrating that von Neumann projection can only be satisfied if
the measurement of ${\bf A}^{(1)}$ is an ideal one (for which
$\lambda_{mm'}= \delta_{mm'}$). In particular, a detector
efficiency smaller than 1 is causing the conditionally prepared
state to deviate from the von Neumann ideal.

\begin{theacknowledgments}
The author likes to thank the Faculty of Applied Physics of
Eindhoven University of Technology, and in particular professor
Thijs Michels, for providing facilities that made this work
possible. Thanks are due also to Andrei Khrennikov for his
invitation to publish the paper in the Proceedings of FPP-5.
\end{theacknowledgments}

\bibliographystyle{aipproc}   % if natbib is available

%\bibliography{vaxjoproc2007}


%\bibliography{methboek}

\end{document}